\documentclass[usenatbib,sn-basic,style=authoryear,
reprint,
superscriptaddress,
nofootinbib,
amsmath,
sn-basic,
amssymb,
aps,
floatfix,
showkeys,
]{sn-jnl}

\usepackage{graphicx,color,xcolor}
\usepackage{amsfonts,amsmath}
\usepackage{amssymb,ulem}
\usepackage{dcolumn}
\usepackage{bm,lipsum}
\usepackage[utf8]{inputenc}
\usepackage{subeqnarray}
\usepackage{array}
\usepackage{epsfig,hyperref}
\usepackage{morefloats}
\usepackage{relsize}
\usepackage{url}

\usepackage{comment}

\usepackage{cleveref}
\usepackage{orcidlink}
\usepackage[]{natbib}
\usepackage{multicol}

\begin{document}

\title{ Macroscopic properties of the XTE J1814-338 as a dark matter admixed strange star}

\author{Luiz L. Lopes} %
\email{llopes@cefetmg.br}

\affil{\orgname{Centro Federal de Educação Tecnológica de Minas Gerais, Campus VIII}, \orgaddress{\street{Av. Imigrantes, 1000}, \city{Varginha}, \postcode{37.02-560}, \state{MG}, \country{Brazil}}}


\date{\today}

\abstract{
In this paper, I discuss the macroscopic properties of the ultracompact object XTE J1814-338, whose inferred mass and radius read $M$ = 1.21 $\pm$ 0.05 $M_\odot$ and R = 7.0 $\pm$ 0.4 km as a dark matter-admixed strange star. By using the neutralino as WIMP dark matter with a fixed Fermi momentum, I calculated the moment of inertia, the gravitational redshift,  the dimensionless tidal parameter,  and the total amount of dark matter for a 1.2$M_\odot$ star.
At the end, I study the role of the neutralino's mass.}

\maketitle


\section{Introduction}

Strange stars are self-bounded compact objects composite by deconfined quarks. The theory of strange stars is based on the so-called Bodmer-Witten conjecture~\citep{Bodmer,Witten},  It assumes that ordinary matter as we know it, composed of protons and neutrons, may only be meta-stable, while the true ground state of strongly interacting matter would therefore consist of the so-called strange quark matter (SQM), which in turn is composed of deconfined up, down and strange quarks. For the SQM hypothesis to be true, the energy per baryon of the deconfined phase (for $p = 0$ and $T = 0$) must be inside the so called stability window~\citep{Lopes2022ApJ}, i.e; the energy per baryon must be lower than the nonstrange infinite baryonic matter, i.e., ($E_{\rm uds}/A < 930$ MeV), while at the same time, the nonstrange matter still needs to have an energy per baryon higher than the one of nonstrange infinite baryonic matter ($E_{\rm ud}/A > 930$ MeV); otherwise, protons and neutrons would decay into $u$ and $d$ quarks. It is possible, therefore, that at least some of the observed pulsars are indeed strange stars instead of ordinary hadronic neutron stars.

 The potential existence of strange stars can explain some mysterious astrophysical phenomena, such as repeating fast radio bursts (FRB). In ~\citet{Geng2021}, it was suggested that the repeating FRB 180916 is produced by the intermittent fractional collapses of the crust of an SS induced by the refilling of materials accreted from its low-mass companion. Future observations will serve as indirect evidence for the strange quark matter hypothesis. 

On the other hand, our current understanding predicts that 27$\%$ of the universe is made of dark matter (DM), 68$\%$ of dark energy (the main component that explains the accelerated expansion of the universe), and only 5$\%$ is luminous matter. Compact objects may capture some amount of DM inside
them in their evolving time due to their immense gravitational potential if non-annihilating DM exist~\citep{Dasgupta2021,Dasgupta2023,Ray2024,Ray2024b}.
Non-annihilating DM can accumulate and thermalize in a small radius, producing changes in all macroscopic properties. While most observed pulsars can be explained as ordinary neutron stars or standard strange stars, ultra-compact objects present a new challenge for theoretical astrophysicists.

The main goal of the present work is to study the nature of the XTE J1814-338 pulsar. The recent work of~\citet{Kini:2024ggu},  has revealed a striking mass (M) and radius (R) measurement of $M = 1.21^{+0.05}_{-0.05}\rm M_\odot$ and $\rm R = 7.0^{+0.4}_{-0.4}km$ at 68\% confidence level (CL). The inner composition of such an ultra-compact object has been a subject of intense research, with some of them speculating that the presence of DM cannot be ruled out. Particularly, some study points to a bosonic DM star with a nuclear matter core \citep{Pitz:2024xvh}, whilst others point to a strange quark star admixed mirrored DM \citep{Yang:2024ycl} and a Fermionic DM admixed neutron star~\citep{lopes2024xte}.  Another possibility is that XTE J1814-338 is a hybrid star, and the small radius being evidence of the existence of twin stars (see ~\citet{laskospatkos2024xte,TsaloukidisPhysRevD.107,SenPhysRevD.106} and the references therein for additional discussions).

The theory of non-annihilating DM-admixed compact objects has already been studied in the literature for neutron stars~\citep{GOLDMAN2013,Panotopoulos:2017idn,Das:2021hnk,Das:2018frc,Lourenco:2021dvh,DeliPRD2019,Lourenco:2022fmf,kumar2024c,issifu2025arxiv}, strange stars~\citep{Odilon2021,Lopes:2023uxi}, and hybrid stars~\citep{Lenzi:2022ypb}. In this work, we revisit this topic, focusing on the macroscopic properties of the XTE J1814-338 pulsar as a DM-admixed strange star. Indeed, such a possibility was already studied in~\citet{Yang:2024ycl}, therefore, it is worth to point out here the differences between the present work and those presented in~\citet{Yang:2024ycl}.

First, the formalism of both phases (quark matter and dark matter) is different. For strange quark matter (SQM), the authors in~\citet{Yang:2024ycl} used the modified bag model, where the effects of gluons were taken into account up to $O(\alpha_s^2)$~\citep{Jaffe1984PRD,WEBER2005}, and used mirror dark matter (MDM), where every Standard Model particle has a mirror counterpart (this means that the mirror SQM (mirror up, down and strange) is governed by the same thermodynamic principles as ordinary quark matter~\citep{Lee1956,Mohapatra_1997,FOOT1991PLB,KHLOPOV2021103824}. Consequently, the EOS for MDM is identical to that of SQM, and the amount of DM is fixed as a fraction of the total star mass.

In the present work, I use the thermodynamical consistent vector MIT bag model, as discussed in detail in~\citet{lopesps1,lopesps2,Carline2023BJP}. In the same sense, instead of mirror dark matter, I use a WIMP dark matter here, specifically the lightest neutralino with a mass of 200 GeV. At the end of the paper, this condition will be relaxed to study the influence of the DM mass. The amount of the DM is fixed by fixing the neutralino Fermi moment, as done in~\citet{Odilon2021,Das:2021hnk,Lopes:2023uxi,Lourenco:2021dvh}. Moreover, I extend the calculation beyond the mass-radius diagram and calculate the gravitational redshift, the moment of inertia (MOI), and the dimensionless tidal parameter. Those calculations are not present in~\citet{Yang:2024ycl}.

\section{Formalism}\label{fm}

\subsection{Strange quark matter}\label{qm}

To model the quark matter, I use the thermodynamically consistent version of the vector MIT bag model, whose Lagrangian density reads:
\begin{eqnarray}
\mathcal{L}_{\rm vMIT} = \bigg\{ \bar{\psi}_q\big[\gamma^\mu(i\partial_\mu - g_{qV} V_\mu) - m_q\big]\psi_q \nonumber \\
- B + \frac{1}{2}m_V^2V^\mu V_\mu  \bigg\}\Theta(\bar{\psi}_q\psi_q) ,  
\label{vMIT}
\end{eqnarray}
where $m_q$ is the mass of the quark $q$ of ﬂavor $u$, $d$ or $s$, $\psi_q$ is the Dirac quark ﬁeld, $B$ is the constant vacuum pressure, and $\Theta(\bar{\psi}_q\psi_q)$ is the Heaviside step function to ensure that the quarks exist only conﬁned to the bag. The parameters utilized in this work are: $m_u = m_d = 4$ MeV, $m_s = 95$ MeV, $B^{1/4} = 140$ MeV, and $G_V = (g_v/m_v)^2 = 0.3$ fm$^2$. Moreover, the obtained EOS is charge-neutral and chemically stable.

It is important to point out that the results are ultimately model-dependent. 
 For instance, by changing $G_V$, we change all the stability window, as discussed in~\citet{lopesps1,lopesps2}. Increasing $G_V$ stiffens the EOS, producing more massive stars, as well as larger stars for a fixed mass value. Decreasing $G_V$ has the opposite effect. On the other hand, fixing $G_V$ but increasing the bag,  produces smaller and less massive stars. Reducing the bag leads to an increase in the maximum mass as well leads to an increase of the radius of a fixed mass value. As pointed out in Fig. 2 in~\citet{lopesps2}, it is possible to reproduce very similar maximum mass with slightly different values of the radius by simultaneously using large values $B$ and $G_V$ or small values $B$ and $G_V$.

The chosen values in the present study, in addition to ensuring that the SQM is inside the stability window and agrees with the Bodmer-Witten conjecture, are able to simultaneously satisfy some astrophysical constraints (as PSR J0740+6620, PSR J0437-4715, and HESS J1731-34). This will become clearer in the next sections.  An additional discussion of the model can be found in~\citet{lopesps1, lopesps2,Carline2023BJP} and the references therein.

\subsection{Dark Matter}\label{dm}

{ Here, I consider the neutralino as a non-annihilating DM.}
The Lagrangian of this fermionic DM has a QHD-like form and reads~\citep{Das:2018frc, Odilon2021, lopes2024xte,Das:2021hnk,Panotopoulos:2017idn,Lopes:2023uxi}:
\begin{eqnarray}
\mathcal{L}_{\rm DM} = \bar{\chi}(i \gamma^\mu \partial_\mu - (m_\chi -g_H h))\chi 
\nonumber \\
+ \frac{1}{2}(\partial^\mu h \partial_\mu h - m_H^2 h^2). \label{FDMEOS}
\end{eqnarray}
Here, we assume a dark fermion represented by the Dirac field $\chi$ that self-interacts through the exchange of the Higgs boson, whose mass is $m_H$ = 125 GeV. The coupling constant is assumed to be $g_H = 0.1$,  which agrees with the constraints in~\citet{Panotopoulos:2017idn, Das:2021hnk}. In this framework, the DM self-interaction is extremely weak, resembling a tenuous, free Fermi gas. The dark matter energy eigenvalue is therefore:

\begin{equation}
 E_{\chi} = \sqrt{m_\chi^{*2} + k^{2}},   
\end{equation}
where $m_\chi^*$ = $m_\chi - g_Hh$ and, in principle, $m_\chi$ = 200 GeV being the mass of the lightest neutralino as done in several studies~\citep{ Das:2021hnk, Lopes:2023uxi,Lenzi:2022ypb,Lourenco:2022fmf}. The mass of the neutralino will be relaxed later. Moreover, following those papers, we use the Fermi momentum to fix the dark matter content, up to $k_F^{DM} = 0.08$ GeV.  As $k << m_\chi$, the pressure of the DM is almost zero.

I also add a term that couples the quark matter with the DM, similarly as it was done for the hadronic case in~\citet{lopes2024xte,Das:2021hnk,Lenzi:2022ypb,Lourenco:2021dvh,Lourenco:2022fmf}.

\begin{equation}
\mathcal{L} = f\frac{m_q}{v} h\bar{\psi_q}\psi_q \label{cc}.
\end{equation}

Within this coupled channel, the effective mass of the quark $q$ now depends on the field $h$. In the same sense, the field $h$ now depends on both the DM scalar density ($n_s^{DM}$) and the quark scalar density $(n_s^q)$~\citep{lopes2024xte,Lenzi:2022ypb,Lourenco:2022fmf}:

\begin{eqnarray}
 m^*_q = m_q  -f\frac{m_q}{v}h ,  \nonumber \\
 h = \frac{g_H}{m_H^2}n_s^{DM} + \frac{f}{m_H^2}\sum_q\frac{m_q}{v}n_s^q , \label{ncc}
\end{eqnarray}
where $v$ = 246 GeV is the Higgs vacuum expectation value, and $f$ = 0.3 as done in~\citet{Lenzi:2022ypb,Lourenco:2022fmf}. The term in Eq.~\ref{cc} is quite weak, being stronger for the $s$-quark compared to the $u$ and $d$ quarks, as expected for the Higgs channel. Also, it is worth pointing out here that changing the coupling parameters of the dark matter ($g_H$ and $f$) will produce only negligible differences in the numerical results.

{ The total energy density and pressure are the sum of the quark and DM components, obtained through a mean-field approximation.
The quark EOS is given by~\citep{lopesps1}:

\begin{eqnarray}
 \epsilon_q = \frac{N_c}{\pi^2}\sum_q \int_0^{k_f} \sqrt{m_q^{*2} + k^2}k^2dk   + \frac{m_V^2V_0^2}{2} + B, \nonumber \\
 p_q = \frac{N_c}{3\pi^2}\sum_q\int_0^{k_f}\frac{k^4 dk}{\sqrt{m_q^{*2} +k^2}} + \frac{m_V^2V_0^2}{2} - B,
\end{eqnarray}
where $N_c = 3$ is the number of colors. The DM EOS reads~\citep{Lenzi:2022ypb}:

\begin{eqnarray}
 \epsilon_{DM} = \frac{1}{\pi^2} \int_0^{k_F^{DM}} \sqrt{m_\chi^{*2} + k^2}k^2dk   + \frac{m_H^2h^2}{2}, \nonumber \\
 p_{DM} = \frac{1}{3\pi^2}\int_0^{k_F^{DM}}\frac{k^4 dk}{\sqrt{m_\chi^{*2} +k^2}} - \frac{m_H^2h^2}{2}.
\end{eqnarray}

}

\section{Numerical Results}\label{ma}

I now present the numerical results of DM admixed strange stars. I began with the equation of state (EOS), i.e; the relation between the pressure and the energy density. Then I use the EOSs as an input to obtain the mass-radius relations by solving the TOV equations~\citep{TOV}. Once the mass-radius relation is obtained, the gravitational redshift, $z$, can be calculated as in~\citet{LopesEPL2021}:

\begin{equation}
 z =  \bigg ( 1 - \frac{2GM}{R} \bigg )^{-1/2} - 1.   
\end{equation}

\begin{figure}[t!]
\centering 
\includegraphics[scale=.48, angle=270]{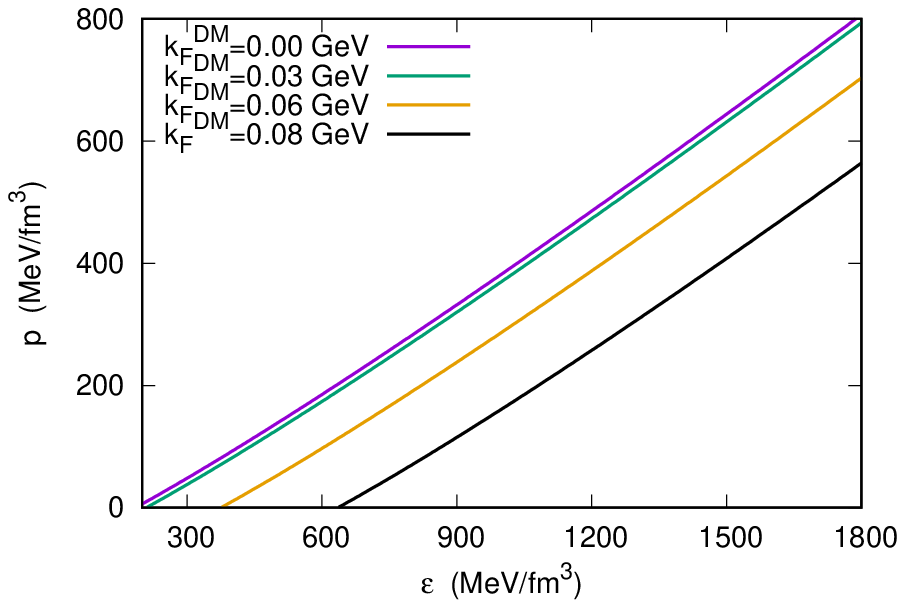} \\
\includegraphics[scale=.48, angle=270]{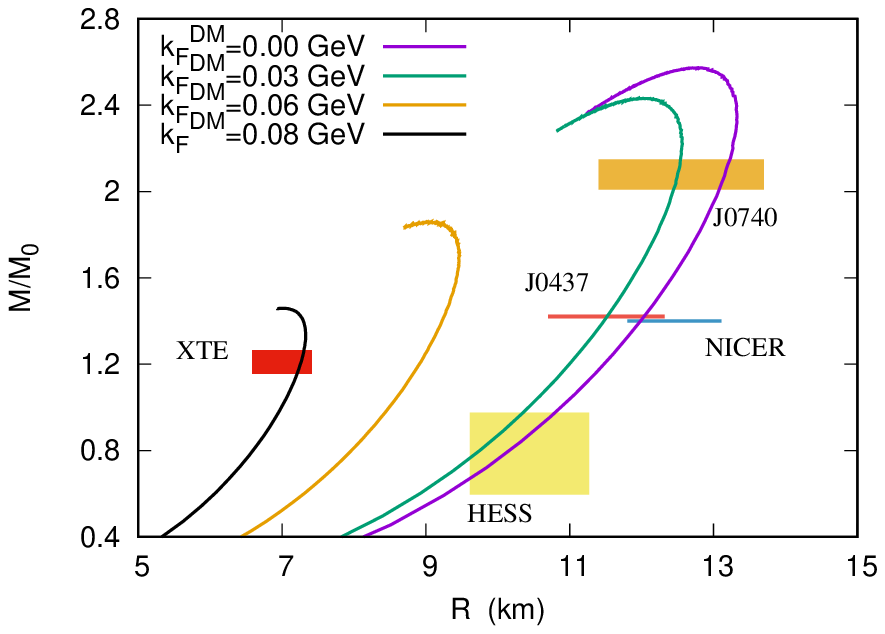} \\
\includegraphics[scale=.48, angle=270]{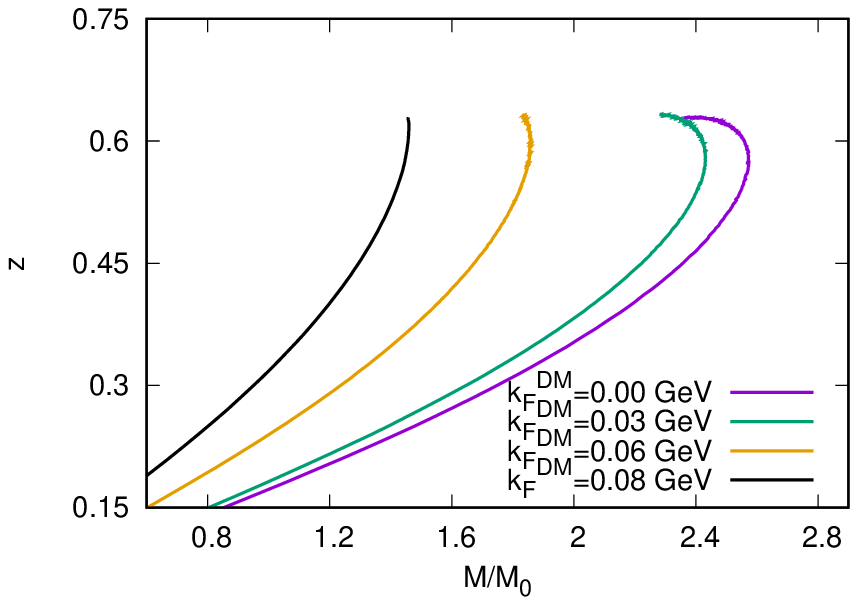}\\
\caption{Strange star's properties for different values of $k_F^{DM}$. Top: the EOSs. Middle: The mass-radius relation together with observational constraints discussed in the text. Bottom: The gravitational redshift.} \label{F1}
\end{figure}

The EOSs are displayed at the top of Fig.~\ref{F1}.
As can be seen, the EOSs are very similar in shape, but as we increase the value of $k_F^{DM}$, the EOS becomes softer. For $k_F^{DM}$ = 0.03 GeV, the EOS is very similar to the EOS without DM. For $k_F^{DM}$ = 0.06 GeV, the EOS is significantly softer, while for $k_F^{DM}$ = 0.08 GeV the softening is extreme.
{ Although the use of $k_F^{DM}$ limited to 0.06 GeV is much more common in the literature ~\citep{Das:2018frc,Lourenco:2021dvh,Lourenco:2022fmf,Lenzi:2022ypb,Lopes:2023uxi}, values up to 0.08 GeV is not new, and was already explored in~\citet{kumar2024c}. The lack of further work using higher $k_F^{DM}$ values can be explained by the fact that the article discussing the XTE J1814-338 object was only published in October 2024~\citep{Kini:2024ggu}. 

Despite the strong effect of the DM, it is worth emphasizing that all the EOSs are causal and mechanically stable, once they obey the Le Chatelier principle, i.e., the quantity $dp/d\epsilon$ lies between 0 and 1.  As the DM is coupled to the quark matter via Eq.~\ref{cc} and \ref{ncc}, the TOV equations are solved in the standard one fluid approach, as done in~\citet{Das:2021hnk,Lopes:2023uxi,Das:2018frc,Lenzi:2022ypb}. All generated stars are stable up to the maximum mass~\citep{Glenbook}.}

In the middle of Fig.~\ref{F1}, I show the mass-radius relations as well as some observational constraints. Maybe the more important constraint is the mass and radius measurement of the  PSR J0740+6620 pulsar, $M = 2.08 \pm 0.07 M_\odot$ and $R =  12.39^{+1.30}_{-0.98}$ km~\citep{Miller2021}, which gives us undoubtedly proof of the existence of ultra-massive neutron stars. For the canonical M = 1.4 M$_{\odot}$ star, we use here the constraint provided by the NICER X-ray telescope, which points to $R_{1.4} = 12.45 \pm 0.65$~\citep{Riley2021},  A standard model (without DM) must be able to satisfy both constraints simultaneously. A recent and very strong constraint is related to the PSR J0437-4715 and was presented in~\citet{Choudhury2024APJL}. The authors claim that this pulsar has a mass and radius of $M = 1.418~\pm~0.037M_\odot$,  and
$R = 11.36^{+0.95}_{-0.63}$ km, which presents a strong constraint
for traditional hadronic stars. Furthermore, besides the XTE J1814-338, with an inferred mass and radius of $M = 1.21 \pm 0.05 M_\odot$ and R = 7.0 $\pm$ 0.4 km~\citep{Kini:2024ggu}, another very compact objects is the HESS J1731-347~\citep{Doroshenko_2022}, whose mass and radius are $M=0.77_{-0.17}^{+0.20}~M_\odot$ and $R = 10.4 _{-0.78}^{+0.86}$ km respectively.

It can be seen from Fig.~\ref{F1} (middle) that a standard strange star (without DM) can simultaneously fulfill the constraints related to the PSR J0740+6620, the PSR J0437-4715, the HESS J1731-347, and the canonical 1.4 $M_\odot$ star. The introduction of DM causes a well-known reduction in both, the radii and maximum masses of the stars~\citep{Das:2018frc,Lenzi:2022ypb,Lourenco:2021dvh}. For $k_F^{DM}$ = 0.03 GeV, the radius of the canonical stars drops below the minimum pointed by the NICER X-ray telescope~\citep{Riley2021}.
{ For $k_F^{DM}$ = 0.06 GeV, we have a more drastic drop in radii and masses, but still not enough to explain the  XTE J1814-338 object, which can only be explained in our model within $k_F^{DM}$ = 0.08 GeV.} I conclude the analysis of the mass-radius relation by pointing out that within $k_F^{DM}$ = 0.08 GeV, a 1.2 $M_\odot$ strange star presents a radius of 7.26 km, while the maximum possible mass reaches 1.46 $M_\odot$.

 Finally, I display at the bottom of Fig.~\ref{F1} the gravitational redshift, $z$, for different values of $k_F^{DM}$. The gravitational redshift is strongly affected by the DM content. For instance, a 1.2 $M_\odot$ strange star without DM has a redshift $z = 0.205$, while within $k_F^{DM}$ = 0.08 GeV, we have $z =0.401$, approximately twice the value. For comparison purposes, ~\citet{LopesEPL2021} showed that the gravitational redshift of the canonical 1.40 $M_\odot$ stars ranges from 0.195 to 0.218 for different radii values. Furthermore, the authors in.~\citet{Sanwal2002APJ} were able to measure and fix the gravitational redshift of the 1E 1207.4-5209 neutron star, pointing out that $z$ must lie in the range $0.12 - 0.23$. As can be seen, the gravitational redshift of the  XTE J1814-338 is far above those presented by the 1E 1207.4-5209 neutron star, indicating that the  XTE J1814-338 must have a completely different nature.
 As a curiosity, it is interesting to point out that despite different values of $k_F^{DM}$ producing very different masses and radii, a similar maximum value of $z$ is found for all values of $k_F^{DM}$, around $z = 0.6$.

 Another interesting feature is the mass fraction of the DM inside the strange star, defined as $f_{DM} = M_{DM}/M$, where:

 \begin{equation}
   M_{DM} = \int_0^R 4\pi \epsilon_{DM} r^2 dr  
 \end{equation}

 For a 1.2 solar mass with a radius of 7.26 km, we obtain $f_{DM} = 0.642$. Therefore, WIMP DM requires a smaller amount of DM compared to MDM, which requires at least a fraction of 0.718~\citep{Yang:2024ycl}.

I now analyze the stars' moment of inertia (MOI) for different values of $k_F^{DM}$. The moment of inertia can be calculated as in~\citet{Glenbook}:

\begin{equation}
  I = \frac{8}{3}\int_0^R r^4 \frac{(\epsilon(r) +p(r))}{\sqrt{1-2GM(r)/r}} \cdot e^{-\nu} dr, \label{inertiam}  
\end{equation}
with

\begin{equation}
  \frac{d \nu}{dr} = G\frac{M(r) +4\pi r^3 p(r)}{r(r -2GM(r))}  .
\end{equation}

\begin{figure}[b!]
\centering 
\includegraphics[scale=.48, angle=270]{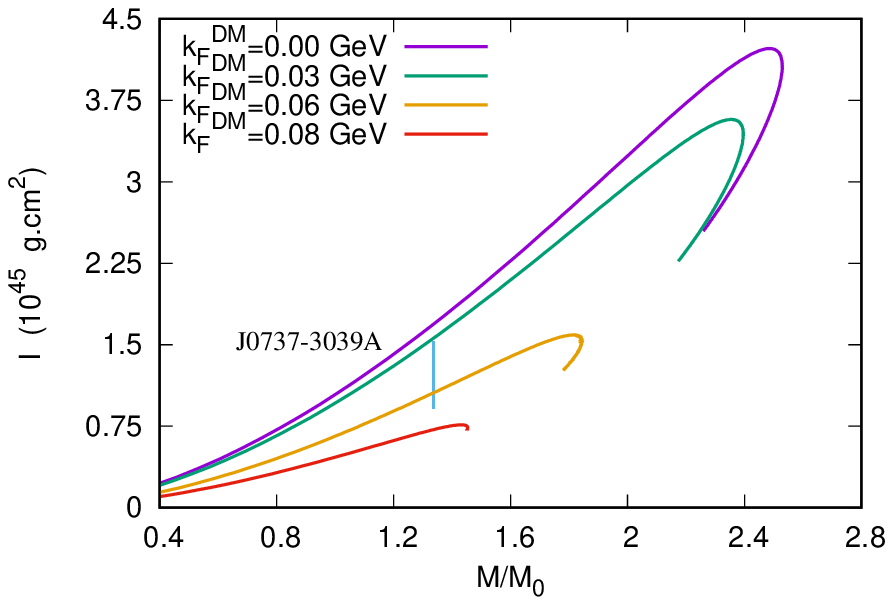} \\
\includegraphics[scale=.48, angle=270]{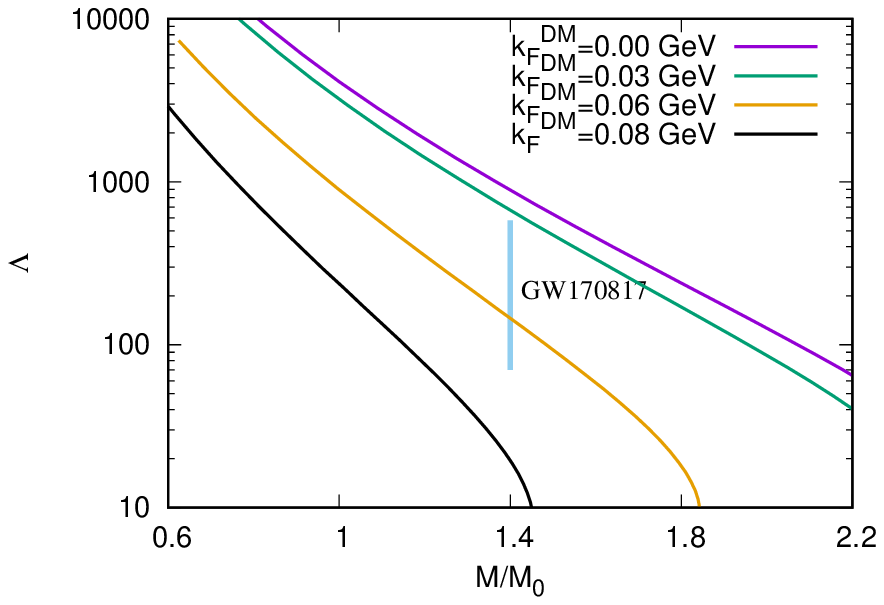} \\
\caption{Strange star's properties for different values of $k_F^{DM}$. Top: MOI. Bottom: Dimensionless tidal parameter.} \label{F2}
\end{figure}

It is important to point out that there is no observational measure of the moment of inertia of neutron stars. Nevertheless, the MOI of the PSR J0737-3039A was estimated as $1.15^{+0.38}_{-0.24}~\times~10^{45}$ g.cm$^2$ based on universal relations analyses~\citep{Landry_2018}, with a well-measured mass of 1.338 $M_\odot$. The results are presented at the top of Fig.~\ref{F2}.

We can see that a standard strange star presents a moment of inertia above the upper limit for the 1.338 $M_\odot$ star, while at the same time for $k_F^{DM}$ = 0.08 GeV it predicts a MOI below the lower limit for the same mass. This reinforces the idea that the XTE J1814-338 must have a different nature when compared to other observed compact objects. About the 1.2 $M_\odot$ star, the presented model predicts a MOI of 0.62 $\times  10^{45}$ g.cm$^2$ for a radius of 7.26 km.

Ultimately, I discuss the dimensionless tidal parameter $\Lambda$.  If we put an extended body in an inhomogeneous external field, it will experience different forces throughout its surface. The result is a tidal interaction. The tidal deformability of a compact object is a single parameter $\lambda$ that quantifies how easily the object is deformed when subjected to an external tidal field. Larger tidal deformability indicates that the object is easily deformed. Conversely, a compact object with a small tidal deformability parameter is more compact and more difficult to deform. The tidal deformability is defined as:
\begin{equation}
 \Lambda~\equiv~\frac{\lambda}{M^5}~\equiv~\frac{2k_2}{3C^5} , \label{tidal}
\end{equation}
where $M$ is the mass of the compact object and $C = GM/R$ is its compactness. The parameter $k_2$ is called the second (order) Love number. Additional discussion about the theory of tidal deformability and the tidal Love numbers is beyond this work's scope and can be found in 
 ~\citet{Lopes2022ApJ,Lopes:2023uxi,Lourenco:2021dvh,Abbott2017, AbbottPRL, Abbott:2018wiz, Flores2020, tidal2010}.  Nevertheless, as pointed out in  ~\citet{Lourenco:2021dvh,tidal2010}, the value of $y_R$ must be corrected when a discontinuity is present.
This is the case of strange stars, once they are self-bounded. Therefore, we must have:
\begin{equation}
 y_R \rightarrow y_R - \frac{4\pi R^3 \Delta\epsilon_S}{3M} , \label{yr}
\end{equation}
where $R$ and $M$ are the star radius and mass, respectively, and $\Delta\epsilon_S$ is the difference between the energy density at the surface ($p=0$) and the star's exterior (which implies $\epsilon=0$). The results are presented in the bottom of the Fig.~\ref{F2} altogether with the constraint coming form the GW170817 event, $70~<~\Lambda_{1.4}~<580$~\citep{AbbottPRL}.

We first notice that the standard strange star has a tidal parameter larger than the upper limit inferred by the GW170817 event, while at the same time, strange stars within $k_F^{DM}$ = 0.08 GeV have a tidal parameter below the lower limit. This fact also corroborates the idea that the  XTE J1814-338 must have a different nature when compared with most observed pulsars. Within $k_F^{DM}$ = 0.08 GeV, the 1.2 $M_\odot$ star has an inferred dimensionless tidal parameter of 78, while the canonical mass has $\Lambda_{1.4}$ = 21.

To finish this part of the study, I present in Tab.~\ref{TL1} all the macroscopic properties of a 1.2$M_\odot$ strange star for different values of $k_F^{DM}$.

 \begin{table}[ht!]
     \centering
\begin{tabular}{c|cccc}
\hline 
$k_F^{DM}$  (GeV) & 0.00 & 0.03 & 0.06 & 0.08\\\hline
$R_{1.2}$ (km) & 11.48 & 10.99 & 8.92 & 7.26 \\
   $z_{1.2}$ & 0.205& 0.216 & 0.291 & 0.401   \\

  $I_{1.2}/10^{45}$ (g.cm$^2$)& 1.42 &  1.30 & 0.89 & 0.62 \\

 $\Lambda_{1.2}$  & 1859 &  1421 & 358 & 78  \\
$f_{DM}$  & 0.00 & 0.118 & 0.503 & 0.642\\
 \hline
\end{tabular}
\caption{ Macroscopic properties of 1.2 $M_\odot$ strange star with different values of $k_F^{DM}$. } 
\label{TL1}
\end{table}

\subsection{Validity of the Results}

{

The validity of the results presented in this work (and, virtually, the validity of the results of any study about the effects of DM on the macroscopic properties of neutron stars~\citep{GOLDMAN2013,Panotopoulos:2017idn,Odilon2021,Das:2021hnk,Lopes:2023uxi,Das:2018frc,Lenzi:2022ypb,Lourenco:2021dvh,DeliPRD2019,Lourenco:2022fmf,kumar2024c}) relies on the assumption that we are dealing with non-annihilating DM. 

However, the chosen dark-matter candidate is neutralino, which is known to potentially annihilate~\citep{BI2006NPB,Kou.PhysRevD2008}. If we allow a self-annihilation DM, the energy released is strong enough to disrupt the entire strange star due to the high DM densities needed to explain the XTE J1814-338 pulsar. For instance, in ~\citet{Kou.PhysRevD2008}, the author has shown that the emissivity of the annihilating DM is given by (Eq. 28):

\begin{equation}
\epsilon = A \times 1.16 \times 10^4~~\mbox{erg}.s^{-1}.\mbox{cm}^{-3}. 
\end{equation}

In this context, A is a constant that parametrizes the local dark matter density near the neutron star, measured in units of 0.3 GeV/cm$^3$. The author studied the emissivity for A varying from 10 to 100.

In this present study, for a $k_{F}^{DM} =0.08$ GeV needed to describe the XTE J1814-338 object, the value of A is of the order of $10^{39}$. Even for a small value of $k_F^{DM} = 0.01$ GeV, we still have A around $10^{36}$.

The luminosity related to these values for a star with a radius of the order of 10 km lies in the range of 10$^{58}$ to 10$^{61}$ erg.s$^{-1}$.

In other words, we are dealing with luminosity above the supernova's scale.

It is clear from the above discussion that the present study is only valid for non-annihilating DM.

Nevertheless, assuming a much lower DM density, other effects can be studied due to the potential DM annihilation. For A = 10 to 100, ~\citet{Kou.PhysRevD2008} shows that DM annihilation affects the temperature of stars older than $10^7$ years, flattening out the temperature at $10^4$ K.

On the other hand, in ~\citet{LeanePRD2021}, the authors compare the standard DM annihilation expected in the halo with DM annihilation in globular clusters, neutron stars, and the Galactic center. It was shown that the signal in celestial bodies can dominate over the halo annihilation rate. As annihilation products may be gamma rays or neutrinos, such signals could potentially be observed by telescopes and observatories like Fermi-LAT or IceCube, respectively.

Additional discussion about DM annihilation is beyond the scope of the present work. The interested reader is referred to see ~\citet{BI2006NPB,Kou.PhysRevD2008,LeanePRD2021} and the references therein.

}

\section{Role of the neutralino's mass}

\begin{figure}[t!]
\centering 
\includegraphics[scale=.52, angle=270]{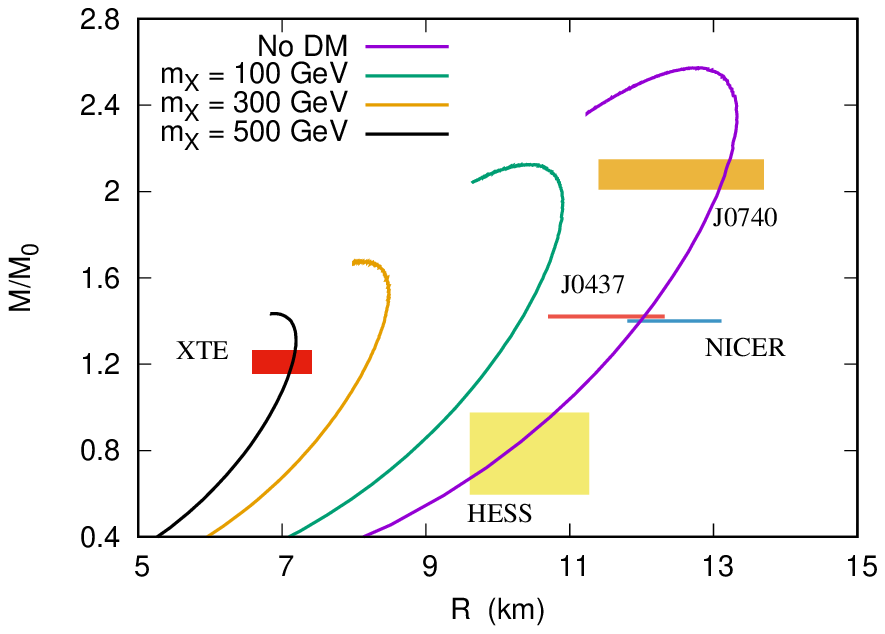} \\
\caption{Strange star's mass-radius relation for different values of $m_X$ with a fixed Fermi momentum $k_F^{DM} =0.06$ GeV.} \label{Ft2}
\end{figure}

So far, I have investigated the effects of DM by changing the Fermi momentum for a fixed neutralino mass. The neutralino mass was chosen equal to 200 GeV as done in other works~\citep{Odilon2021,Das:2021hnk,Lopes:2023uxi,Lourenco:2021dvh}, while the Fermi momentum was increased till be able to reach the mass and radius range of the XTE J1814-338 pulsar. Here, I invert the situation and study the effect of the neutralino mass for a fixed $k_F^{DM} = 0.06$ GeV, which is the upper limit of several studies~\citep{Lopes:2023uxi, Das:2018frc,Lenzi:2022ypb, Lourenco:2021dvh, Lourenco:2022fmf}. The mass values used are 100 GeV, 300 GeV, and 500 GeV. The results are displayed in Fig.~\ref{Ft2}.

As can be seen, for a fixed Fermi momentum, increasing the neutralino mass compresses the stars similarly to increasing the Fermi momentum for a fixed value of $m_\chi$. To reproduce the XTE J1814-338 pulsar, we need $m_\chi$ = 500 GeV for $k_F^{DM}$ = 0.06 GeV. For a 1.2 $M_\odot$ pulsar, I found $R_{1.2}$ = 7.15 km and $f_{DM}$ = 0.648. These values are very close to the ones obtained with $k_F^{DM}$ = 0.08 GeV and $m_\chi$ = 200 GeV, as seen in Tab.~\ref{TL1}. The maximum mass obtained is also similar, 1.43 $M_\odot$ vs 1.46 $M_\odot$. 

In order to avoid this section being redundant to the previous ones, here I only discuss the results relative to $m_\chi$ = 500 GeV, which is able to explain the XTE J1814-338 pulsar, and compare the results with $k_F^{DM}$ = 0.08 GeV with $m_\chi$ = 200 GeV. 
In relation to the redshift, I obtain $z_{1.2}$ = 0.412, which is a little higher than the one obtained with $k_F^{DM}$ = 0.08 GeV. This was expected since the radius was slightly lower (7.15 vs 7.26 km). The same is true for MOI, $I_{1.2}$ = 0.60 $\times$ 10$^{45}$ g.cm$^2$ and the dimensionless tidal parater, $\Lambda_{1.2} = 71$.

The explanation for the great similarity between these results lies in the almost equal value of the DM energy density: we have $\epsilon_{DM}$ = 475 MeV/fm$^{3}$ for a $m_\chi$ = 500 GeV, with $k_F^{DM}$ = 0.06 GeV, and $\epsilon_{DM}$ = 450 MeV/fm$^{3}$ for   $m_\chi$ = 200 GeV, with $k_F^{DM}$ = 0.08 GeV. The slightly superior value of energy density in the first case also explains the slightly lower radius.  It is also worth pointing out that the pressures are significantly different, around one order of magnitude. However, both are very small (around 10$^{-3}$ MeV/fm$^3$ for $m_\chi$ = 200 GeV and 10$^{-4}$ MeV/fm$^3$ for $m_\chi$ = 500 GeV).

\section{Final Remarks}\label{fr}

In this work, I investigate the macroscopic properties of the ultra-compact object XTE J1814-338 pulsar, assuming it to be a DM-admixed strange star. I first studied the total amount of DM necessary to compress a 1.2$M_\odot$ to a radius in the range of 7.0 $\pm$ 0.4 km. I found that a high value for the DM Fermi momentum, $k_F^{DM}$ = 0.08 GeV, is needed. Such a high value implies that the fraction of DM inside of the star, $f_{DM}$, corresponds to 64.2$\%$ of the star's total mass. Although a considerable value, this is still lower than an MDM-admixed strange star, which requires at least 71.8$\%$~\citep{Yang:2024ycl}.

It can be noticed from the TOV solution in Fig.~\ref{F1} that it is very unlikely that the XTE J1814-338 object presents the same nature as other pulsars, such as the PSR J0437-4715 and the PSR J0740+6620. The results of the gravitational redshift reinforce this idea.

Analyzing the MOI and the dimensionless parameter of the tidal $\Lambda$, we see that while standard strange stars present values above the upper limits, in the case of strange stars with DM compatible with the XTE J1814-338 object, the calculated values are below the lower limits. Again, this suggests a different nature relative to different observed pulsars. If the mass and radius inferred for the XTE J1814-338 object presented in ~\cite{Kini:2024ggu} are confirmed, this will result in new and strong indirect evidence of the existence of DM particles.

 As neutralinos can potentially self-annihilate, I check if the results can still be valid if we are dealing with self-annihilating DM. Due to the enormous density reached in the interior of compact objects,  the results are only valid if we strictly consider non-annihilating DM. The  XTE J1814-338 cannot be adequately explained when annihilation processes are considered.

Ultimately, I remade the analysis and fixed the Fermi moment at 0.06 GeV while changing the neutralino mass. I found that a mass of around 500 GeV is needed to explain the XTE J1814-338 pulsar. The other macroscopic quantities are very similar to those produced with $m_\chi$ = 200 GeV and $k_F^{DM}$ = 0.08 GeV.

This can be explained by similar values of the DM energy density.

\textbf{Acknowledgements:} L.L.L.  was partially supported by CNPq (Brazil)
under Grant No 305347/2024-1 and Grant No 409029/2021-1.

Ethics declaration: not applicable

\bibliography{abref}

\end{document}